\documentclass[iop,revtex4]{emulateapj}
\usepackage{lscape} 
\usepackage{natbib}

\usepackage{rotating}
\usepackage{amsmath}
\newcommand \kms{km~$\rm{s}^{-1}$}

\newfont{\rten}{cmr10} 
\def \arcdeg{\hbox{$^\circ$}}

\begin{document}

\title{Origin of 1I/'Oumuamua. II. An ejected exo-Oort cloud object?}

\author{Amaya Moro-Mart\'{\i}n$^{1}$}

\altaffiltext{1}{Space Telescope Science Institute, 3700 San Martin Dr., Baltimore, MD 21218. email: amaya@stsci.edu}

\begin{abstract}
1I/'Oumuamua is the first detected interstellar interloper. We test the hypothesis that it is representative of a background population of exo-Oort cloud objects ejected under the effect of post-main sequence mass loss and stellar encounters. We do this by comparing the cumulative number density of interstellar objects inferred from the detection of 1I/'Oumuamua to that expected from these two clearing processes. We consider the 0.08--8 M$_{\odot}$ mass range, take into account the dependencies with stellar mass, Galactocentric distance, and evolutionary state, and consider a wide range of size distributions for the ejected objects. Our conclusion is that 1I/'Oumuamua is likely not representative of this background population, even though there are large uncertainties in the masses and size distributions of the exo-Oort Clouds.
We discuss whether the number density of free-floating, planetary-mass objects derived from gravitational microlensing surveys could be used as a discriminating measurement regarding 1I/'Oumuamua's origin (given their potential common origin). We conclude that this is challenged by the mass limitation of the surveys and the resulting uncertainty of the mass distribution of the free floaters. The detection of interlopers may be one of the few observational constraints of the small end of this population, with the caveat that, as we conclude here and in Moro-Mart\'{\i}n (\citeyear{2018ApJ...866..131M}), in the case of 1I/'Oumuamua, it might not be appropriate to assume this object is representative of an isotropic background population, which makes the derivation of a number density very challenging.
\end{abstract}


\section{Introduction}
\label{Introduction}

1I/'Oumuamua, first detected by PanSTARRS (Williams \citeyear{Williams2017}), shows a clearly hyperbolic orbit ($e$ = 1.197, $a$ =  -1.290 au, $q$ = 0.254 au, and $i$ = 122.6) and high pre-encounter velocity (26.22 \kms, Mamajek \citeyear{2017RNAAS...1a..21M}). Its origin has triggered great interest because it is the first interstellar interloper to have ever been detected. Based on its photometry (showing a 2--2.5 magnitude variability) and assumed albedo, it is estimated to have an elongated shape (with an axis ratio ranging from 3 to 10) and an effective radius in the range of 55--130 m (Jewitt et al. \citeyear{2017ApJ...850L..36J}, Banninster et al. \citeyear{2017ApJ...851L..38B}, Meech et al. \citeyear{2017Natur.552..378M}, Drahus et al. \citeyear{2018NatAs...2..407D}, Bolin et al. \citeyear{2018ApJ...852L...2B}, Fraser et al. \citeyear{2018NatAs...2..383F}), the uncertainties arising from its unknown shape and albedo. 1I/'Oumuamua's tumbling state is consistent with a bulk density of $\sim$ 1 g/cm$^3$ (Drahus et al. \citeyear{2018NatAs...2..407D}). 

Even though cometary activity was not observed (Jewitt et al. \citeyear{2017ApJ...850L..36J}, Meech et al. \citeyear{2017Natur.552..378M}), there has been a recent detection of non-gravitational acceleration in the outbound orbit of 1I/'Oumuamua that has been interpreted by some authors as evidence of outgassing and therefore of a cometary composition (Micheli et al. \citeyear{2018Natur.559..223M}). Under this scenario, the lack of activity would be attributed to the presence of a thin insulating mantle (Micheli et al. \citeyear{2018Natur.559..223M}). This interpretation is challenged by the absence of a coma, in spite of the implied high mass-loss rate (with comas being observed in comets with significantly lower mass-loss rates -- David Jewitt, private communication), and by the expectation that the implied outgassing torques would have spinned-up the object in a timescale of few days, leading to its breakup (Rafikov \citeyear{2018ApJ...867L..17R}). However, a recent study by Seligman et al. (in preparation) of the rotational dynamics of 1I/'Oumuamua found that its observed stable rotation period could be consistent with the reported non-gravitation acceleration if caused by outgassing from a jet launched from the sub-stellar point of maximal solar irradiation.
	
As summarized in Moro-Mart\'{\i}n (\citeyear{2018ApJ...866..131M}), different authors have studied the cumulative number density of 1I/'Oumuamua-like objects, using a range of values for the detection volume and survey lifetime, and have calculated from these estimates what the contribution per star would need to be to account for the inferred cumulative number density. They then discussed whether these estimates would agree with expectations, given a range of potential origins (Gaidos et al. \citeyear{2017RNAAS...1a..13G}; Laughlin \& Batygin \citeyear{2017RNAAS...1a..43L}; Trilling et al., \citeyear{2017ApJ...850L..38T}; Do et al. \citeyear{2018ApJ...855L..10D}; Feng \& Jones \citeyear{2018ApJ...852L..27F}; Rafikov \citeyear{2018ApJ...861...35R}; Raymond et al. \citeyear{2018MNRAS.476.3031R}; Portegies-Zwart et al. \citeyear{2018MNRAS.479L..17P}, Moro-Mart\'{\i}n \citeyear{2018ApJ...866..131M}).

The two main potential sources considered in these studies are the following. 

\begin{itemize}
\item {\it Planetesimal disk origin}. Planetesimal ejection is a natural product of the planetesimal/planet formation processes and is efficient for a wide range of planetary architectures (Raymond et al. 2018b and references therein). Because the majority of the ejected material would have formed outside the snowline in their parent systems, this has led to the idea that the interstellar space must be filled with planetesimals of a predominantly icy composition (Moro-Mart\'{\i}n et al. \citeyear{2009ApJ...704..733M}). In  Moro-Mart\'{\i}n (\citeyear{2018ApJ...866..131M}), we explored this hypothesis and concluded that 1I/'Oumuamua is unlikely representative of such a population of isotropically distributed objects, even when considering a wide range of size distributions, favoring the scenario proposed by Gaidos et al. (\citeyear{2017RNAAS...1a..13G}) and Gaidos (\citeyear{2018MNRAS.477.5692G}) that 1I/'Oumuamua originated from the planetesimal disk of a young nearby star, in which case the ejected bodies would not be isotropically distributed. That origin would be in agreement with several indicators of 1I/'Oumuamua's youth: its color, not as red as the ultra-red bodies in the outer solar system, thought to be reddened by space weathering from cosmic rays and ISM plasma (Jewitt et al. \citeyear{2017ApJ...850L..36J}); its tumbling state, indicating an age younger than its inferred damping time scale of $\sim$ 1 Gyr (Drahus et al. \citeyear{2018NatAs...2..407D}); and most significantly its kinematic properties, yielding an age of $\ll$1 Gyr, based on its low velocity with respect to the Local Standard of Rest (LSR), 3--10 \kms, indicating the object has not been subjected to significant dynamical heating by passing stars, clouds or spiral arms (Gaidos et al. \citeyear{2017RNAAS...1a..13G}; Mamajek \citeyear{2017RNAAS...1a..21M}, Do et al. \citeyear{2018ApJ...855L..10D}). 
\item {\it White dwarf or supernova origin}.  The lack of cometary activity observed in 1I/'Oumuamua has been considered by some authors to indicate a refractory nature. This has led to studies that have explored whether 1I/'Oumuamua could have been ejected from a white dwarf system in a tidal disruption event  (Rafikov \citeyear{2018ApJ...861...35R}) or via direct ejection (Hansen \& Zuckerman \citeyear{2017RNAAS...1a..55H}), or as the result of the large and sudden mass loss of a supernova explosion.
\end{itemize}

In this paper, we address another potential origin for 1I/'Oumuamua: exo-Oort clouds.  If the cometary nature of 1I/'Oumuamua were to be confirmed, this origin would be of interest because of the predominantly icy composition of these objects, expected to have been formed beyond the snowline of their parent systems, although there is evidence that some members of the solar system's Oort Cloud may have originated in the asteroid belt (Meech et al. \citeyear{2016SciA....2E0038M}). 

In the solar system, the Oort cloud is thought to have formed due to the interplay of planetary scattering and external forces: the forming giant planets scattered the planetesimals in this region out to large distances where they were subject to external influences, like the slowly changing gravitational potential of the cluster, the Galactic tides, and the stellar flybys, with different models favoring different perturbers.  These external influences would have caused the perihelion distances of the scattered planetesimals to be lifted to distances $>>$ 10 au,  where the planetesimals were no longer subject to further scattering events but were also safe from complete ejection and thus remained weakly bound to the solar system, forming the Oort cloud (see for example Brasser et al. \citeyear{2012Icar..217....1B}). Some authors argue that the Oort cloud formed while the Sun was in its birth cluster. Under this scenario, the main perturbers would be the stars and gas in the cluster. These models, however, fail to account for the circularization of the orbits due to the cluster gas (that would impede the comets to be scattered out into the Oort cloud, Brasser et al. \citeyear{2010A&A...516A..72B}), and for the stripping of the outer parts of the Oort cloud ($\ge$ 3 $ \cdot~$~10$^{4}$) by the cluster gravitational potential and neighboring stars.  To account for these caveats, other authors argued that the Oort cloud formed during the late dynamical instability of the solar system, about 0.5 Gyr after it formed. The caveat of these latter models is that this process is not sufficiently efficient (by an order of magnitude) to account for the estimated number of bodies in the Oort cloud (derived from the flux of long-period comets) based on the estimated mass in planetesimals that would have remained $\sim$0.5 Gyr after the solar system formed, i.e. after most of the protoplanetary disk was dispersed (Brasser et al. \citeyear{2010A&A...516A..72B}). 

Even though the formation of the solar system's Oort cloud has still many unknowns, we can expect exo-Oort clouds to form around other stars as the result of the interplay of planetary scattering and external forces that would lead to the lifting of the periastrons of bodies initially orbiting closer to the star (Wyatt et al. \citeyear{2017MNRAS.464.3385W}). Indirect evidence of the presence of a reservoir of comets around other stars are the debris disk systems. There is also evidence that some of these exocomets have been scattered into the inner regions of these system, as suggested by the observation of variable absorption gas features in several of these debris disks (Kiefer et al. \citeyear{2014Natur.514..462K}, Welsh \& Montgomery \citeyear{2015AdAst2015E..26W}), and by the dips in the lightcurve of some Kepler sources (Boyajian et al. \citeyear {2016MNRAS.457.3988B}). 

The reason why we are interested in these exo-Oorts clouds as a potential source of interlopers like 1I/'Oumuamua is because dynamical models show that, over the lifetime of their parent stars, these weakly bound objects are subjected to ejection due to Galactic tides, post-main sequence mass loss, and encounters with other stars or with giant molecular clouds (Veras et al. \citeyear{2011MNRAS.417.2104V},  \citeyear{2012MNRAS.422.1648V}, \citeyear{2014MNRAS.437.1127V}). For  example, for the solar system, Hanse et al. (\citeyear{2018MNRAS.473.5432H}) indicated that over the Sun's main sequence, the Oort cloud will lose 25-65\% of its mass due mainly to stellar encounters, with a second stage of Oort cloud clearing to be triggered by the onset of mass loss as the Sun enters the post-main sequence stage (Veras et al. \citeyear{2012MNRAS.422.1648V}). These ejected objects will contribute to the population of free-floating material and this contribution is expected to be more significant in the Galactic bulge than in the disk or the halo of the Galaxy (due to the more frequent stellar encounters in the former), and in the oldest regions than in the youngest regions (due to the timescale associated to the clearing processes, Veras et al. \citeyear{2014MNRAS.437.1127V}). 

The efficiency of exo-Oort cloud formation is unknown. In this study we explore the scenario in which exo-Oort clouds are common and can lead to a background population of ejected exo-Oort cloud objects. 

The capture by the solar system of one of these ejected exo-Oort cloud objects today is highly unlikely due to their expected high relative velocity with respect to the Sun, but may have been possible when the solar system was still embedded in its maternal birth cluster (Levison et al. \citeyear{2010DDA....41.0201L}; Belbruno \citeyear{2012AsBio..12..754B}), 
with the higher transfer efficiencies being enabled by the lower relative stellar velocities, an order of magnitude lower than today. There is therefore the possibility that we have already observed, or will be able to observe, one of these captured objects, but its origin beyond the solar system will likely remain uncertain. The interest of exo-Oort cloud objects possibly crossing the solar system today is precisely because, given their high relative velocity with respect to the Sun, they would be clearly distinguishable from other solar system objects, as it has been the case of 1I/'Oumuamua, that happened to pass very close to the Earth and now is on its way out of the solar system (Jewitt et al. \citeyear{2017ApJ...850L..36J}). 

The goal of this study is to assess  whether 1I/'Oumuamua could be representative of an isotropically distributed  population of ejected exo-Oort cloud objects. Do et al. (\citeyear{2018ApJ...855L..10D}) addressed this question for the case of intermediate-mass stars (2--8 M$_{\odot}$), where the ejection is triggered by post-main sequence mass loss. They concluded that this process may account for the inferred number density of interstellar objects based on 1I/'Oumuamua's detection, favoring the scenario in which 1I/'Oumuamua's is a freed exo-Oort cloud object. Our calculation improves upon that in Do et al. (\citeyear{2018ApJ...855L..10D}) because: (1) it takes into account that the contribution from each star to the population of interstellar objects depends on the mass and Galactocentric distance of the star, and its evolutionary state; (2) it considers a wider range of stellar masses (0.08--8 M$_{\odot}$); (3) it includes the contribution from main sequence stars due to stellar encounters; and (4) it considers a wide range of size distributions for the ejected objects. 

In Section \ref{OverallUpperLimit}, we carry out a back-of-the-envelope calculation that anticipates the main conclusions of this study. This is followed by a more detailed discussion of the expected contribution from stars in the 0.08--8 M$_{\odot}$ mass range to the population of interstellar objects, a calculation that takes into account the effect of the different perturbing forces and its dependence on the stellar mass, Galactocentric distance, and evolutionary state (Sections \ref{Clearing} and \ref{Contribution}), and the effect of considering a wide range of possible size distributions for the ejected objects (Sections \ref{Contribution} and \ref{Discussion}). Our conclusions are summarized in Section \ref{Conclusion}. 

\section{Back-of-the-envelope estimate of the number of ejected objects}
\label{OverallUpperLimit}

\subsection{Exo-Oort cloud assumptions: number of objects}
\label{number}

For this back-of-the-envelope calculation we need an estimate of the number of objects that each star could potentially contribute. 

Given the lack of observational constraints, our assumption regarding the number of objects in the exo-Oort clouds is based on solar system observations. In the solar system, long-period comets are thought to originate in the Oort cloud and launched into the inner solar system due to perturbations by the Galactic tide, or by encounters with stars or giant molecular clouds, together with subsequent perturbations by the planets. From the flux of long-period comets, one can estimate the number of objects in this cloud to be  10$^{11}$--10$^{12}$, with (1--5) $ \cdot~$~10$^{11}$ being the most likely value (Brasser \& Morbidelli \citeyear{2013Icar..225...40B} and references therein). As Brasser et al. (\citeyear{2010A&A...516A..72B}) pointed out, this estimate is uncertain because the flux of long-period comets with $q<$ 4 au is only 2--3 per year, so an error of only 1 comet per year can significantly affect the result. In this study, we assume that the Oort cloud contains $\sim$ 10$^{12}$ objects with diameters $\geqslant$ 2.3 km. 

Following Hanse et al. (\citeyear{2018MNRAS.473.5432H}), for a given exo-Oort  cloud we assume that its number of bodies larger than 2.3 km is similar to that of the solar system's Oort cloud scaled to the mass of the parent star, in our case 10$^{12} \left({{\it M_{*}} \over {M_{\odot}}}\right)$. We base this assumption on the following considerations. 

The models of Brasser et al. (\citeyear{2010A&A...516A..72B}) found that the Oort cloud formation efficiency is similar at a wide range of Galactocentric distances. It will, however, depend on the environment and the planetary architecture of the system. Regarding the latter, Figure 1 in Wyatt et al. (\citeyear{2017MNRAS.464.3385W}) shows that the parameter space to form an Oort cloud is quite restricted and in the solar system is populated by Uranus and Neptune. Their figure also shows that the parameter space to have a population of ejected bodies (that never become part of the Oort Cloud) is well populated by known exoplanets, Jupiter in the case of the solar system, so it may be that the presence of this gas giant resulted in a relatively depleted Oort Cloud. With this in mind, on the one hand, given the restricted parameter space to form an exo-Oort cloud, assuming that all stars harbor exo-Oort clouds is likely an overestimate. On the other hand, given that the parameter space for ejected bodies is well populated by known exoplanets, and that these objects would contribute to the background population under consideration (regardless of whether or not they become trapped in an Oort cloud before being completely ejected from the system), assuming all planet-bearing stars contribute about 10$^{12} \left({{\it M_{*}} \over {M_{\odot}}}\right)$ objects  might be a reasonable order-of-magnitude estimate. We will go one step further and assume that all stars contribute, irrespective of whether or not they are planet hosts. This latter assumption likely makes our back-of-the-envelope estimate an upper limit. 

The above estimate assumes that the size distribution of the objects ejected by other systems is similar to that of the solar system, which is of course unknown. Because the ejection processes are independent of size, another approach would be to estimate the mass that could be ejected per star. That was the approach taken in Moro-Mart\'{\i}n (\citeyear{2018ApJ...866..131M}) when exploring protoplanetary disks as a potential origin of 1I/'Oumuamua, and the observational constraint in that case was the available mass of solids in protoplanetary disks. In this study, we are working with a complementary observational constraint: the number of long-period comets and inferred population of the solar system's Oort cloud.

\subsection{Resulting upper limit of the number of ejected objects with sizes equal or larger than 1I/'Oumuamua's}
\label{upperlimit}

Given the above considerations, we can already calculate an upper limit to the number of ejected objects with sizes equal or larger than 1I/'Oumuamua's contributed by stars in the 1--8 M$_{\odot}$ mass range. This would be  be given by 
\begin{equation}
\begin{split}
N_{\rm ejected}^{\rm max} = 
\int\limits_{1M_{\odot}}^{8M_{\odot}} \xi({\it M_{*}})\left({\rm 0.16 \over \rm 2.3}\right)^{\rm -2.5}\rm 10^{12} \left({{\it M_{*}} \over {M_{\odot}}}\right)d{\it M_{*}}.
\end{split}
\label{eqUpperLimit}
\end{equation}

In Equation \eqref{eqUpperLimit}, $\xi(M_*)$ is the initial mass function for which we adopt the one derived from Kroupa et al. (\citeyear{1993MNRAS.262..545K}) that finds that the number density of  stars per pc$^3$, out to $\sim$ 130 pc from the Sun, within the mid-plane of the galaxy, and with stellar masses between {\it M$_{*}$} and {\it M$_{*}$+dM$_{*}$} (in units of M$_{\odot}$), is given by\footnote{Equation \eqref{eq_stellar_density} corresponds to an initial mass function derived based on the present-day mass function, assuming no significant stellar evolution for low-mass stars, and using Scalo's (\citeyear{1986FCPh...11....1S}) initial mass function for higher masses (Kroupa et al. \citeyear{1993MNRAS.262..545K}).} 

\begin{equation}
\begin{split}
n(M_*) = \xi(M_*)dM_*~~~~{\rm with,}~~~~~~~~~~~~\\ 
\xi(M_*) = 0.035 M_*^{-1.3}~~~~{\rm if~0.08} \le M_* < 0.5\\
\xi(M_*) = 0.019 M_*^{-2.2}~~~~~{\rm if~0.5} \le M_* < 1.0\\
\xi(M_*) = 0.019 M_*^{-2.7}~~~~{\rm if~1.0} \le M_* < 100.  
\end{split}
\label{eq_stellar_density}
\end{equation}
 
The term $\left({\rm 0.16 \over \rm 2.3}\right)^{\rm -2.5}\rm 10^{12} \left({{\it M_{*}} \over {M_{\odot}}}\right)$ in Equation \eqref{eqUpperLimit} corresponds to the cumulative number of objects with sizes equal or larger than 1I/'Oumuamua's, assuming an equilibrium power law size distribution with index $q$ = 3.5, $n(r) \propto r^{-3.5}$, and adopting for 1I/'Oumuamua the intermediate effective radius of 80 m from Drahus et al. (\citeyear{2018NatAs...2..407D}).  

As we mentioned above (in Section \ref{number}), this estimate would likely be an upper limit because it assumes all stars contribute, irrespective of whether or not they are planet hosts. It also assumes that if exo-Oort clouds form they eventually get entirely depleted due to the clearing processes described in Section \ref{Introduction}. The resulting estimate is $N_{\rm ejected}^{\rm max}$ = 1.6$ \cdot~$10$^{13}$  pc$^{-3}$.

\subsection{Comparison of the back-of-the-envelope upper limit to the inferred population of interstellar objects based on 1I/'Oumuamua's detection}
\label{Comparison_back-of-envelope}

We can now compare the above estimate to that inferred from the detection of 1I/'Oumuamua. For the latter, we adopt the number density distribution derived by Do et al. (\citeyear{2018ApJ...855L..10D}) of $N_{\rm r \geqslant R}$ =  0.21 au$^{-3}$ $\sim$ 2$ \cdot~10^{15}$ pc$^{-3}$, an estimate that assumes that the objects are isotropically distributed, and adopts for 1I/'Oumuamua an absolute magnitute of H=22.1, a nominal phase function with slope parameter $G$ = 0.15, a velocity at infinity, $v_{\infty}$ = 26 \kms, and a detection frequency of 1 per 3.5 years (assumed to be the survey's lifetime). Do et al. (\citeyear{2018ApJ...855L..10D}) noted that the presence of a size distribution leads to an approximately  40\% uncertainty, while the existence of inefficiencies in the detection process leads to a cumulative number density (4/3-3/2) $\times$ their estimated value. 

For comparison, other authors\footnote{For reference, these other estimates  assume a range of survey times of 1--2 yrs (Jewitt et al. \citeyear{2017ApJ...850L..36J}), 5 yrs (Portegeis-Zwart et al. \citeyear{2018MNRAS.479L..17P}), 7 yrs (Gaidos et al. \citeyear{2017RNAAS...1a..13G}), and 20 yrs (Feng et al. \citeyear{2018ApJ...852L..27F}), compared to the 3.5 years assumed by Do et al. (\citeyear{2018ApJ...855L..10D}). They also assume  a small range of dark albedos and absolute magnitudes that result in an object radius of 55 m (Jewitt et al. \citeyear{2017ApJ...850L..36J}), 60 m (Fraser et al. \citeyear{2018NatAs...2..383F}), 100 m (Portegeis-Zwart et al. \citeyear{2018MNRAS.479L..17P}), 115 m (Gaidos et al. \citeyear{2017RNAAS...1a..13G}), and 50 m (Feng et al. \citeyear{2018ApJ...852L..27F}), compared to 80 m assumed by Do et al. (\citeyear{2018ApJ...855L..10D}).} estimate that the number density is 0.1 au$^{-3}$ = 8$ \cdot~10^{14}$ pc$^{-3}$ (Jewitt et al. \citeyear{2017ApJ...850L..36J}, Fraser et al. \citeyear{2018NatAs...2..383F}), 0.012--0.087 au$^{-3}$ = 1--7$ \cdot~10^{14}$ pc$^{-3}$ (Portegies-Zwart et al. \citeyear{2018MNRAS.479L..17P}), 0.012 au$^{-3}$ = 1$ \cdot~10^{14}$ pc$^{-3}$ (Gaidos et al. \citeyear{2017RNAAS...1a..13G}), and $>$ 0.006 au$^{-3}$ = 4.8$ \cdot~10^{13}$ pc$^{-3}$ (lower limit from Feng et al. \citeyear{2018ApJ...852L..27F}). We adopt the cumulative number density estimate from Do et al. (\citeyear{2018ApJ...855L..10D}) because, unlike the other studies, it is derived based on a careful calculation of the PanSTARRS detection volume. However, as mentioned above, we note that there are uncertainties in their inferred value. 

Taking in consideration that our estimate is likely an upper limit, and noting it is approximately two orders of magnitude smaller than the inferred value based on 1I/'Oumuamua's detection, this already leads to the main conclusion of this study: 1I/'Oumuamua is unlikely representative of a background population of ejected exo-Oort cloud objects from 1--8 M$_{\odot}$ stars. This conclusion is only as strong as the assumptions about the masses and size distributions of exo-Oort Clouds and assumes that the size distribution of the ejected population is similar to that of the solar system and that it follows and equilibrium power law with index $q$ = 3.5, $n(r) \propto r^{-3.5}$.  For further discussion on this result, the reader is referred to Sections \ref{Discussion} and \ref{Conclusion}. 

In Sections \ref{Clearing} and \ref{Contribution} we go beyond this back-of-the-envelope calculation and discuss in more detail the expected contribution from stars in the 0.08--8 M$_{\odot}$ mass range to the population of interstellar objects, taking into account the effect of the different perturbing forces and its dependence on the stellar mass, Galactocentric distance, and evolutionary state. In Sections \ref{Contribution} and \ref{Discussion}, we also consider a wide range of possible size distributions for the ejected objects, based on solar system observations and on accretion and collisional models. These more detailed calculations reinforce the main conclusion of the study.
 
\section{Exo-Oort cloud clearing processes}
\label{Clearing}
 
Veras et al. (\citeyear{2011MNRAS.417.2104V},  \citeyear{2012MNRAS.422.1648V}, \citeyear{2014MNRAS.437.1127V}) studied how post-main sequence mass-loss, stellar encounters, and the Galactic tide affect the dynamical evolution of an orbiting body, and the prospects for its ejection from the system during the entire stellar lifetime (main sequence, post-main sequence, and white dwarf stages). 

\subsection{Exo-Oort cloud assumptions: extent}
\label{Extent}

The effect of the perturbing forces mentioned above will depend on the semimajor axis of the orbiting body. Also in this case, given the lack of observational constraints, our assumption regarding the extent of the exo-Oort clouds is based on solar system observations. In the solar system, as we mentioned above, long-period comets are thought to originate in the Oort cloud. Based on long-period comet observations, it is estimated that the solar system's Oort cloud has an isotropic distribution with perihelion $q$ $\gtrsim$ 32 au and inner and outer semimajor axes of $a_{\rm min}^{\rm OC}$ $\sim$ 3 $ \cdot~$~10$^{3}$ and $a_{\rm max}^{\rm OC}$ $\sim$ 10$^{5}$ au, respectively. 

Following Hanse et al. (\citeyear{2018MNRAS.473.5432H}), we scale the inner and outer edge of a given exo-Oort cloud to the Hill radius of its parent star in the Galactic potential, so that its inner and outer semimajor axes are  
\begin{equation}
\begin{split}
a_{\rm min}^{\rm exo-OC} \sim a_{\rm min}^{\rm OC} \cdot~{R_{*}\over R_{0\odot}} \left({M_{*}\over M_{\odot}}{M_{\rm G}(R_{0\odot})\over M_{\rm G}(R_{*})}\right),\\
a_{\rm max}^{\rm exo-OC} \sim a_{\rm max}^{\rm OC} \cdot~{R_{*}\over R_{0\odot}} \left({M_{*}\over M_{\odot}}{M_{\rm G}(R_{0\odot})\over M_{\rm G}(R_{*})}\right), 
\end{split}
\label{OCradius}
\end{equation}
where $R_{*}$ and $R_{0\odot}$ are the Galactocentric radii of the parent star and the Sun, respectively, and $M_{\rm G}(R_{*})$ and $M_{\rm G}(R_{0\odot})$ are the Galactic masses contained within those radii. Following this scaling law, Table 1 lists the inner and outer exo-Oort cloud radii for 1 $M_{\odot}$ and 8 $M_{\odot}$ stars at three different Galactocentric distances (4 kpc, 8.5 kpc, and 12 kpc), together with the z-component of their Hill ellipsoid (the latter taken from  Figure 3 in Veras et al. \citeyear{2014MNRAS.437.1127V}). Given these assumed exo-Oort cloud radial extents, in Sections \ref{Clearing_PMS}, \ref{Clearing_enc}, and \ref{Clearing_tides}, we address the effect of the perturbing forces under consideration.
 
\subsection{Post-main sequence mass loss}
\label{Clearing_PMS}

Regarding the perturbation of post-main sequence mass loss, Veras et al. (\citeyear{2011MNRAS.417.2104V},  \citeyear{2012MNRAS.422.1648V}, \citeyear{2014MNRAS.437.1127V}) improved upon previous calculations by removing the traditional "adiabatic" approximation\footnote{The adiabatic approximation is not valid in this case because  we one cannot assume that the mass-loss timescale is much larger than the object's orbital timescale (Veras et al. \citeyear{2011MNRAS.417.2104V}).} and solving instead the full variable-mass, two-body problem, assuming isotropic mass-loss and adopting a realistic multiphasic non-linear, mass-loss prescription (important because the strength and duration of the post-main sequence mass-loss strongly affect the results in the solar-mass range). Depending on the stellar properties, that determine the strength and timescale of the post-main sequence mass-loss, they found that this mass loss can result in the ejection of Oort cloud objects. 

For stars in the 0.8--0.9 M$_{\odot}$ mass range, Veras et al. (\citeyear{2011MNRAS.417.2104V}) found that, even though they lose half of their mass in their post-main sequence stage, this mass loss is not strong enough to eject bodies from their exo-Oort clouds. 
For stars in the 1--8 M$_{\odot}$ mass range, they found that their mass loss is strong enough to guarantee the ejection of a fraction of the Oort cloud-like bodies. 
For more massive stars, 8--20 M$_{\odot}$, the sudden mass loss experienced during the supernova explosion would also trigger the ejection of orbiting bodies at Oort-cloud distances. However, given the uncertainty of whether these latter stars might harbor exo-Oorts (among other things because of the exo-Oort cloud formation timescale), in this study we do not consider the contribution of potential exo-Oort clouds around stars in this high-mass range. 
To conclude: under our assumptions, and based on Veras' results, the dominant contribution of post-main sequence mass loss to the ejected population of exo-Oort cloud object would come from 1--8 M$_{\odot}$ stars. 

For stars in the 1--8 M$_{\odot}$ mass range, Figure 3 in Veras et al. (\citeyear{2014MNRAS.437.1127V}) shows the regions where the dynamical evolution of their orbiting bodies would be under the influence of post-main sequence mass-loss (black lines), together with the two other processes under consideration: Galactic tide (brown lines), and stellar encounters (pink lines). They calculated these regions by comparing the orbital timescale of the orbiting body to the timescales of these three different dynamical processes. In their figure, the degree to which post-main sequence mass loss and the Galactic tide affect the dynamics of the orbiting objects are measured by the nondimensional index $\psi$, defined as the ratio of the orbital timescale of the object to the timescale of the process under consideration. Veras et al. (\citeyear{2014MNRAS.437.1127V}) points out that $\psi$ = 0.02 is the standard value; smaller values of $\psi$ (0.001) would correspond to adiabaticity transitions for objects which are sensitive to small changes in orbital behavior (like comets with high eccentricity); while larger values of $\psi$ (0.7) would correspond to higher thresholds for nonadiabatic orbital change. Depending on the value of $\psi$, the system becomes nonadiabiatic to different extents and from this information one can estimate whether bodies orbiting at a given semimajor axis range could be subject to ejection\footnote{Note, however, that the fate of a given object depends strongly on its orbit, location at the time of the onset of the post-main sequence mass loss, and strength and duration of the mass loss (Veras et al. \citeyear{2014MNRAS.437.1127V}).} (Veras et al. \citeyear{2014MNRAS.437.1127V}). 

To see how post-main sequence mass loss and the two other potential exo-Oort cloud clearing processes may affect the exo-Oort clouds, and considering the unknown orbital evolution of the Sun in the Galaxy, we extract three vertical slides from Figure 3 in Veras et al. (\citeyear{2014MNRAS.437.1127V}), corresponding to three different Galactocentric distances: 4 kpc, 8.5 kpc, and 12 kpc. This assumes the stars are in circular orbits around the Galactic center and takes into account that stars in this range of Galactocentric distances, encompassing the possible migration of the Sun in the Galaxy, could have contributed to the population of ejected exo-Oort cloud objects encountering the solar system. The information extracted from these slides is summarized in Table 1 that lists, among other quantities, the adiabaticity boundaries corresponding to post-main sequence mass-loss of stars in the 1--8 M$_{\odot}$ mass range for different vales of the $\psi$ index (0.02 in solid line; 0.001 in dash-dotted line; 0.7 in dashed line). A body orbiting at a semimajor axis beyond these boundaries would be subject to become dynamically unstable. Ejection, however, is not guaranteed, as collision with the star is in some cases a possible outcome (Veras et al. \citeyear{2014MNRAS.437.1127V}). The location of the adiabaticity boundary corresponding to post-main sequence mass loss is independent on Galactocentric distance and is closer to the star in the case of more massive stars, making their exo-Oort cloud objects more prone to escape.


\renewcommand\thetable{1}
\LongTables
\begin{deluxetable*}{lllllll}
\tablewidth{0pc}
\tablecaption{Regimes of influence of the different exo-Oort cloud clearing processes (in AU)\tablenotemark{a}}
\tablehead{
\colhead{} &
\colhead{1 M$\odot$} &
\colhead{1 M$\odot$} &
\colhead{1 M$\odot$} &
\colhead{8 M$\odot$} &
\colhead{8 M$\odot$} &
\colhead{8 M$\odot$}\\
\colhead{} &
\colhead{4 kpc} &
\colhead{8.5 kpc} &
\colhead{12 kpc} &
\colhead{4 kpc} &
\colhead{8.5 kpc} &
\colhead{12 kpc}}
\startdata
Inner boundary\tablenotemark{b} ($a_{\rm min}^{\rm exo-OC}$) 					& 1.7$ \cdot~$10$^{3}$			& 3$ \cdot~$10$^{3}$ 				& 3.9$ \cdot~$10$^{3}$		& 3.4$ \cdot~$10$^{3}$		& 6$ \cdot~$10$^{3}$ 				& 7.7$ \cdot~$10$^{3}$\\
Outer boundary\tablenotemark{b} ($a_{\rm max}^{\rm exo-OC}$) 					& 5.7$ \cdot~$10$^{4}$			& 1$ \cdot~$10$^{5}$ 				& 1.3$ \cdot~$10$^{5}$		& 1.1$ \cdot~$10$^{5}$		& 2$ \cdot~$10$^{5}$ 				& 2.6$ \cdot~$10$^{5}$\\
Hill radius (z-comp.)													& 9.4$ \cdot~$10$^{4}$			& 1.5$ \cdot~$10$^{5}$ 			& 2.1$ \cdot~$10$^{5}$			& 1.9$ \cdot~$10$^{5}$ 			& 3.1$ \cdot~$10$^{5}$ 			& 4.1$ \cdot~$10$^{5}$\\
Galactic tide ($\psi$=0.02) 											& 3.3$ \cdot~$10$^{4} (79.9\%)$	& 5.5$ \cdot~$10$^{4}$ (83\%) 	& 7.8$ \cdot~$10$^{4}$ (77.5\%)	& 1.6$ \cdot~$10$^{5}$ (0\%)		& 1.1$ \cdot~$10$^{5}$ (83\%)	& 1.6$ \cdot~$10$^{5}$ (77.5\%)\\
Galactic tide ($\psi$=0.001) 										& 1.2$ \cdot~$10$^{4}$ (98.9\%)	& 2.1$ \cdot~$10$^{4}$ (99\%)	& 2.9$ \cdot~$10$^{4}$ (98.8\%)	& 2.5$ \cdot~$10$^{4}$ (98.9\%)	& 4.2$ \cdot~$10$^{4}$ (99\%)	& 5.8$ \cdot~$10$^{4}$ (98.8\%)\\
Galactic tide ($\psi$=0.7) 											& 1.2$ \cdot~$10$^{5}$ (0\%)		& 1.9$ \cdot~$10$^{5}$ (0\%)		& 2.6$ \cdot~$10$^{5}$ (0\%)		& 2.3$ \cdot~$10$^{5}$ (0\%)		& 3.8$ \cdot~$10$^{5}$ (0\%)		& 5.1$ \cdot~$10$^{5}$ (0\%)\\
Post-MS mass loss ($\psi$=0.02) 									& 2.5$ \cdot~$10$^{2}$ (100\%) 	& 2.5$ \cdot~$10$^{2}$ (100\%)	& 2.5$ \cdot~$10$^{2}$ (100\%)	& 2.7$ \cdot~$10$^{1}$ (100\%)	& 2.7$ \cdot~$10$^{1}$ (100\%)	& 2.7$ \cdot~$10$^{1}$ (100\%)					\\
Post-MS mass loss ($\psi$=0.001) 									& 3.4$ \cdot~$10$^{1}$ (100\%)	& 3.4$ \cdot~$10$^{1}$ (100\%)	& 3.4$ \cdot~$10$^{1}$ (100\%)	& 4.1 (100\%)						& 4.1~~~~~~(100\%)				& 4.1 (100\%)					\\
Post-MS mass loss ($\psi$=0.7) 									& 2.8$ \cdot~$10$^{3}$ (100\%)	& 2.8$ \cdot~$10$^{3}$ (100\%)	& 2.8$ \cdot~$10$^{3}$ (100\%)	& 3.2$ \cdot~$10$^{2}$ (100\%)	& 3.2$ \cdot~$10$^{2}$ (100\%)	& 3.2$ \cdot~$10$^{2}$ (100\%)	\\
Stellar encounters\tablenotemark{c} (MS) 												& 79 (100\%)						& 2.5$ \cdot~$10$^{2}$ (100\%)	& 5.9$ \cdot~$10$^{2}$ (100\%)	& 1.4$ \cdot~$10$^{3}$ (100\%)	& 4.3$ \cdot~$10$^{3}$ (100\%)	& 1.0$ \cdot~$10$^{4}$ (100\%)\\
Stellar encounters\tablenotemark{c} (TPAGB)											& 5.5$ \cdot~$10$^{3}$ (99.9\%)	& 1.7$ \cdot~$10$^{4}$ (99.5\%) & 4.1$ \cdot~$10$^{4}$ (97\%)	& 1.3$ \cdot~$10$^{4}$ (99.9\%)	& 4.2$ \cdot~$10$^{4}$ (99.1\%)	& 9.9$ \cdot~$10$^{4}$ (94.4\%)\\
\tablenotetext{a}{Objects orbiting at semimajor axis larger than the listed values would be subject to perturbations due to the process under consideration that could lead to escape. Shown in parenthesis is the fraction of the corresponding exo-Oort cloud that would lie inside a given regime of influence.}
\tablenotetext{b}{Calculated from equation \eqref{OCradius} assuming $M_{\rm G}$(4 kpc) = 2.1 $ \cdot~$~10$^{11}$ M$_{\odot}$, $M_{\rm G}$(8.5 kpc) = 3.7 $ \cdot~$~10$^{11}$ M$_{\odot}$, and $M_{\rm G}$(12 kpc) = 4.8 $ \cdot~$~10$^{11}$ M$_{\odot}$, in agreement with the Galactic gravitational potential derived by Watkins et al. (\citeyear{2018arXiv180411348W}) corresponding to the sum of the galactic nucleus, bulge, disk and halo.}
\tablenotetext{c}{Closest stellar distances derived by Veras et al. (\citeyear{2014MNRAS.437.1127V}) using a Galaxy model that closely reproduces observations of all the local stellar kinematics. Note that these closest stellar distances have not been derived directly from the initial mass function described in Section \ref{eq_stellar_density}.}

\end{deluxetable*}
 
\subsection{Stellar encounters}
\label{Clearing_enc}

In the absence of dynamical simulations that take into consideration the details of the stellar encounters, Veras et al. (\citeyear{2014MNRAS.437.1127V}) studied the regime in which this effect might be important by comparing the semimajor axis of the orbiting body to the closest expected encounter distance over a given timescale, where for the latter they consider the main sequence lifetime (labeled MS) and the thermally pulsating AGB lifetime (labeled TPAGB). Table 1 lists these values, corresponding to 1 $M_{\odot}$ and 8 $M_{\odot}$ stars at Galactocentric distances of 4 kpc, 8.5 kpc, and 12 kpc. These results indicate that the closest expected encounter distance has a strong Galactocentric distance dependance and that close flybys may be common during both stages of stellar evolution, implying that exo-Oort cloud erosion due to stellar flybys is an important clearing process that needs to be considered. For the Sun, for example, this distance extends into the scattered disk/inner solar system region during the main sequence stage. 

Regarding the strength of the effect (measured by the index $\psi$ in the other two processes considered), in the case of stellar encounters, Veras et al. (\citeyear{2014MNRAS.437.1127V}) pointed out that it is not possible to predict it without detailed dynamical simulations because it depends strongly on the unknown orientation of the stellar collision. Hanse et al. (\citeyear{2018MNRAS.473.5432H}) carried out detailed dynamical simulations to estimate Oort cloud erosion and the possibility of capturing exo-comets during close encounters with other stars harboring similar Oort clouds. They assumed three different possible orbits of the Sun in the Galaxy (migrating inward, migrating outward and not migrating), leading to different stellar encounter sequences. They found that, over the lifetime of the Sun, the Oort cloud will lose 25\%--65\% of its mass mainly due to stellar encounters with stars in the 0.9--8.1 M$_{\odot}$ mass range (the Galactic tide not playing a significant role); they also found that the transfer of comets happens only during the extremely rare, relatively slow ($\le$ 0.5 \kms), close ($\le$ 10$^{5}$ au) flybys, and are often lost shortly after the encounter, due to the Galactic tide or consecutive encounters with other stars, leading to a fraction of only $\sim$ 10$^{-5}$--10$^{-4}$ of captured exo-comets in the Oort cloud (c.f. Levison et al. \citeyear{2010DDA....41.0201L}). 
 
 \subsection{Galactic Tides}
\label{Clearing_tides}

Table 1 also shows the location of the adiabaticity boundaries corresponding to the Galactic tide. They often lie inside the exo-Oort clouds being considered, indicating this process also contributes to their erosion. However, given that the timescale for the Galactic tide to affect the dynamical evolution of an Oort cloud-like object is of the order of a Gyr (Veras et al. \citeyear{2011MNRAS.417.2104V}), whereas stellar encounters and post-main sequence mass loss operate in a much faster timescale, and given that the boundaries corresponding to these two other processes are significantly closer to the star\footnote{As it can be seen in Table 1, while the star is experiencing post-main sequence mass loss, the adiabaticity boundary corresponding to this process  is more than an order of magnitude smaller than that corresponding to the Galactic tide for all $\psi$ values considered, so that the former process dominates over the latter during this stage of stellar evolution.}, the latter two are expected to dominate the clearing.
  
\subsection{Conclusions regarding the dominant clearing processes}

As Table 1 indicates, the two processes expected to dominate the exo-Oort clouds clearing are: (1) post-main sequence mass loss, for the stars in the 1--8 M$_{\odot}$ mass range that have reached this stage of stellar evolution; and (2) stellar encounters, for the stars that are still on their main sequence. It becomes critical, therefore, to estimate how many stars in the mass range considered have reached the end of their main sequence. In Section \ref{Contribution}, we estimate this value and their expected contribution to the population of ejected exo-Oort cloud objects from the two main clearing processes mentioned above. 

\section{Contribution to the population of ejected exo-Oort cloud objects}
\label{Contribution}

\subsection{Post-main sequence mass loss}
\label{Contribution_PMS}

Based on the discussion in Section \ref{Clearing}, we calculate the contribution of post-main sequence mass loss of stars in the 1--8 M$_{\odot}$ mass range to the cumulative number density of ejected exo-Oort cloud objects with sizes equal or larger than 1I/'Oumuamua's, $N_{\rm exo-OC}^{\rm PMS}$, using this expression:
\begin{equation}
\begin{split}
N_{\rm exo-OC}^{\rm PMS} = \int\limits_{1M_{\odot}}^{8M_{\odot}} \xi({\it M_{*}})\it f_{\rm PMS}({\it M_{*}})\\
\times~\left({N_{r \geqslant R_1}\over N_{r \geqslant R_2}}\right)  \rm 10^{12} \left({{\it M_{*}} \over {M_{\odot}}}\right){\it f_{\rm eject}^{\rm PMS}}({\it M_{*}})d{\it M_{*}},
\end{split}
\label{NPMS}
\end{equation}
where the different terms are described below. 

\subsubsection{Number density of stars that have reached the post-main sequence stage: $\xi({\it M_{*}})\it f_{\rm PMS}({\it M_{*}})$}
\label{SfPMS}
The term $\xi(M_*)$ is the initial mass function described in Equation \eqref{eq_stellar_density} (from Kroupa et al. \citeyear{1993MNRAS.262..545K}), while the term $f_{\rm PMS}({\it M_{*}})$ corresponds to the fraction of stars that have reached the post-main sequence stage at the current time. To calculate this fraction, following Kroupa et al. (\citeyear{1993MNRAS.262..545K}), we assume a constant star formation rate over the age of the Galactic disk (for which we adopt the value of $t_{disk}$ = 1.2$ \cdot~$10$^{10}$ yr). We also assume that the stars arrive at the main sequence at time $t_{0}$ and that their main sequence lifetime is $t_{ms}$, with
\begin{equation}
\begin{split}
log(t_0(M_*)) = -5.380 M_* + 9.0322~~~~{\rm if}~M_* \le 0.15\\
log(t_0(M_*)) = -0.9398 M_* + 8.3543~~~~{\rm if}~M_* > 0.15,
\end{split}
\label{eq_t0}
\end{equation}
and 
\begin{equation}
t_{MS}(M_*) = 10^6\left({2550+669 M_*^{2.5} + M_*^{4.5} \over 0.0327 M_*^{1.5} + 0.346 M_*^{4.5}}\right), 
\label{eq_tm}
\end{equation}
with $t_0(M_*)$ and $t_{MS}(M_*)$ in years (from Kroupa et al. \citeyear{1993MNRAS.262..545K}). 

Under the above assumptions, the fraction of stars that have reached the post-main sequence stage at the current time is approximated by 
\begin{equation}
f_{\rm PMS}(M_*) = {t_{disk}-(t_0(M_*)+t_{MS}(M_*))\over t_{disk}}.
\label{fPMS}
\end{equation}

\subsubsection{Cumulative number of bodies with sizes equal or larger than 1I/'Oumuamua's: $\left({N_{r \geqslant R_1}\over N_{r \geqslant R_2}}\right)\rm 10^{12} \left({{\it M_{*}} \over {M_{\odot}}}\right)$}
\label{cumOu}
As discussed in Section \ref{upperlimit}, for a given exo-Oort cloud, we assume that the number of bodies with diameter larger than 2.3 km is similar to that of the solar system's Oort cloud scaled to the mass of the parent star, 10$^{12} \left({{\it M_{*}} \over {M_{\odot}}}\right)$. Multiplying by the factor ${\bf \left({N_{r \geqslant R_1}\over N_{r \geqslant R_2}}\right)}$, with $R_1$ = 0.08 km and $R_2$ = 1.15 km, converts this number to the cumulative number of objects with sizes equal or larger than 1I/'Oumuamua's (adopting for 1I/'Oumuamua the intermediate effective radius of 80 m from Drahus et al. \citeyear{2018NatAs...2..407D}). To calculate this factor, following Moro-Mart\'{\i}n (\citeyear{2018ApJ...866..131M}), we assume that the size distribution can be approximated as the following broken, based on solar system observations and on accretion and collisional models:

\begin{equation}
\begin{split}
n(r) \propto r^{-q_1}~{\rm if}~r_{min}<r<r_b\\
n(r) \propto r^{-q_2}~{\rm if}~r_b<r<r_{max}\\
q_1 = {\rm 2.0,~2.5,~3.0,~3.5}\\
q_2 = {\rm 3,~3.5,~4,~4.5,~5}\\
r_b = {\rm 3~km, 30~km,~90~km}\\
r_{max} \approx~{\rm1000 km}.\\
\end{split}
\label{2s-sizedist}
\end{equation}

This leads to
\begin{equation}
\begin{split}
{N_{r \geqslant R_1}\over N_{r \geqslant R_2}} = \\
{{1 \over -q_1+1}\left[r_b^{-q_1+1}-R_1^{-q_1+1}\right]+{r_b^{q_2-q_1} \over -q_2+1}\left[r_{max}^{-q_2+1}-r_b^{-q_2+1}\right] \over {1 \over -q_1+1}\left[r_b^{-q_1+1}-R_2^{-q_1+1}\right]+{r_b^{q_2-q_1} \over -q_2+1}\left[r_{max}^{-q_2+1}-r_b^{-q_2+1}\right]},
\end{split}
\label{factor}
\end{equation}
with $q_1$, $q_2$, and $r_b$  adopting the values listed above, and with $R_1$ = 0.08 km and $R_2$ = 1.15 km. 

The equilibrium power-law size distribution with index $q$ = 3.5, $n(r) \propto r^{-3.5}$, is a specific case of this broken power-law where $q_1 = q_2 = 3.5$. For this particular case, ${N_{r \geqslant R_1}\over N_{r \geqslant R_2}} = \left({\rm 0.16 \over \rm 2.3}\right)^{\rm -2.5}$. The values listed in Table 2 correspond to this specific case, while Figures \ref{2s} and \ref{2sl} show the results when adopting the broken, power-law size distribution in Equation \eqref{2s-sizedist}.



\subsubsection{Efficiency of ejection of exo-OC bodies due to post-main sequence mass loss: ${\it f_{\rm eject}^{\rm PMS}}({\it M_{*}})$}
\label{fejectpms}
As mentioned in Section \ref{Clearing_PMS}, the adiabaticity boundaries listed in Table 1 indicate that exo-Oort cloud objects are subject to ejection due to post-main sequence mass loss. This, however, does not guarantee that all the objects beyond these boundaries will be ejected, as collision with the star is in some cases a possible outcome (Veras et al. \citeyear{2014MNRAS.437.1127V}). To estimate the efficiency of ejection, we adopt the results in Veras et al. (\citeyear{2014MNRAS.437.1127V}) regarding the range of semimajor axes where objects are guaranteed to escape due to post-main sequence mass loss, and the fraction of the entire main sequence Hill ellipsoid that contain these orbits. These latter fractions are approximately 51.5\%, 74.7\%, 82.7\%, 85.3\%, 87.3\%, 87.8\%, 88.3\% and 89.0\%, for stellar masses of 1 $M_{\odot}$, 2 $M_{\odot}$, 3 $M_{\odot}$, 4$M_{\odot}$, 5 $M_{\odot}$, 6 $M_{\odot}$, 7 $M_{\odot}$, and 8 $M_{\odot}$, respectively (from Veras et al. \citeyear{2014MNRAS.437.1127V}). From these values, we calculate the stellar mass dependency of ${\it f_{\rm eject}^{\rm PMS}}({\it M_{*}})$. 

This is only an approximation because our assumed exo-Oorts do not occupy the entire Hill ellipsoid. In addition, even though Veras' models adopt a realistic multiphasic, non-linear, mass-loss prescription for stars in the solar-mass range, Veras et al. (\citeyear{2012MNRAS.422.1648V}) noted that the results regarding the fraction of Oort cloud bodies that may be ejected in this mass range are also uncertain because of the unknown mass-loss variability during this stage of stellar evolution. 

\subsection{Stellar encounters}
\label{Contribution_enc}

The contribution of stellar encounters of main-sequence stars in the 1--8 M$_{\odot}$ mass range to the cumulative number density of ejected exo-Oort cloud objects with sizes equal or larger than 1I/'Oumuamua's, $N_{\rm exo-OC}^{\rm enc}$, is given by:
\begin{equation}
\begin{split}
N_{\rm exo-OC}^{\rm enc}  = \int\limits_{1M_{\odot}}^{8M_{\odot}} \xi({\it M_{*}})(\rm 1-\it f_{\rm PMS}({\it M_{*}}))\\
\times~\left({N_{r \geqslant R_1}\over N_{r \geqslant R_2}}\right) \rm 10^{12} \left({{\it M_{*}} \over {M_{\odot}}}\right){\it f_{\rm eject}^{\rm enc}}({\it M_{*}})d{\it M_{*}}, 
\end{split}
\label{Nenc}
\end{equation}
where the different terms are described below.  

\subsubsection{Number density of stars on the main sequence: $\xi({\it M_{*}})(1-\it f_{\rm PMS}({\it M_{*}}))$}
\label{SfMS}
Following the discussion in Section \ref{SfPMS}, the number density of stars on the main sequence is given by $\xi({\it M_{*}})(1-\it f_{\rm PMS}({\it M_{*}}))$.  Here, we ignore the contribution from stellar encounters of stars in the post-main sequence phase (even though the closest stellar encounter distance, listed in Table 1, are well within the exo-Oort clouds considered) and in the white dwarf stage. In Section \ref{OverallUpperLimit}, we considered these contributions in the calculation of the back-of-the-envelope estimate. 

\subsubsection{Cumulative number of bodies with sizes equal or larger than 1I/'Oumuamua's: ${\bf \left({N_{r \geqslant R_1}\over N_{r \geqslant R_2}}\right)}\rm 10^{12} \left({{\it M_{*}} \over {M_{\odot}}}\right)$}
See Section \ref{cumOu} for a discussion of this term. 

\subsubsection{Efficiency of ejection of exo-Oort cloud bodies due to stellar encounters: ${\it f_{\rm eject}^{\rm enc}}({\it M_{*}})$}
\label{fejectenc}
We adopt the models by Hanse et al. (\citeyear{2018MNRAS.473.5432H}) for the solar system's Oort cloud that estimate that, over the lifetime of the Sun, and considering a range of its possible orbits in the Galaxy, the Oort cloud will lose 25\%--65\% of its mass mainly due to stellar encounters with stars in the 0.9--8.1 M$_{\odot}$ mass range. As mentioned in Section \ref{Clearing_enc}, this estimate remains uncertain because of the unknown details regarding the stellar encounter series and the fact that the ejection process is dominated by a small number of the strongest encounters, with the  unknown orientation of the stellar collision strongly determining the outcome. However, Hanse et al. (\citeyear{2018MNRAS.473.5432H}) concluded that the considerable Oort cloud erosion due to stellar encounters is a robust result. We therefore adopt the value of ${\it f_{\rm eject}^{\rm enc}}({\it M_{*}})$ = 0.5 for the ejection efficiency. This estimate would be an upper limit because the ejection efficiencies calculated by Hanse et al. (\citeyear{2018MNRAS.473.5432H}) correspond to the entire main sequence lifetime of the Sun and, by definition, the stars considered here would not have reached the end of their main sequence (the younger stars contributing a smaller fraction of ejected objects than the older stars). Note also that these simulations are for a 1 M$_{\odot}$ and we are assuming a similar fraction for other stellar masses. This is not necessarily accurate because the binding energy of an exo-Oort cloud body will depend on the mass of the central star, and so does the exo-Oort cloud location.

\subsection{Resulting contributions}
\label{ResultingContributons}
\begin{figure}
\begin{center}
\includegraphics[width=8cm]{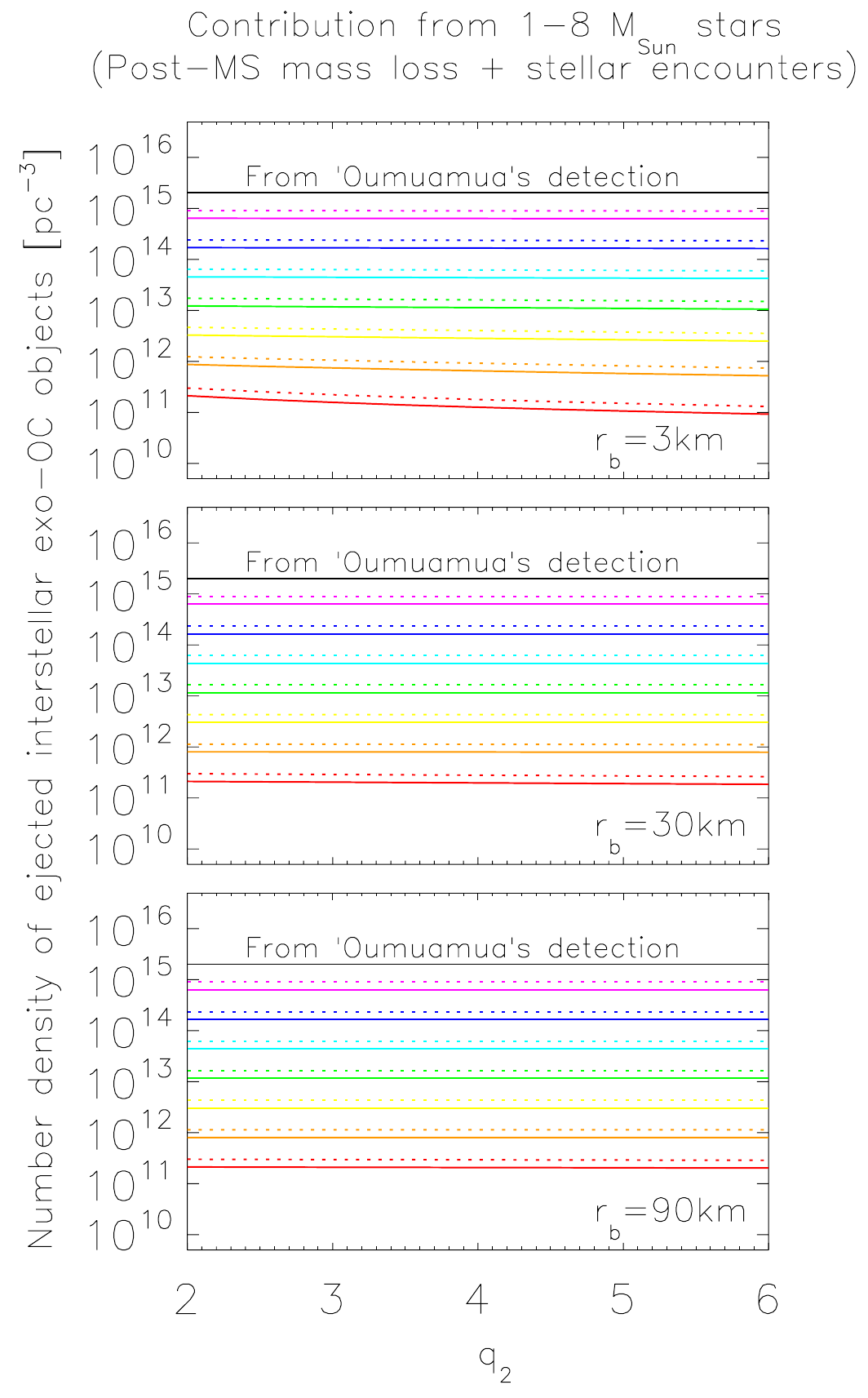}
\end{center}
\caption{Total contribution from 1--8 M$_{\odot}$ stars to the population of ejected exo-Oort cloud objects due to the two exo-Oort cloud clearing processes discussed in Section \ref{Contribution}, adding up the results of Equation \ref{NPMS} (due to post-main sequence mass loss), and Equation \ref{Nenc} (due to stellar encounters). The different colors correspond to different $q_1$ values of the power law size distribution described in Section \ref{2s-sizedist}, with $q_{\rm 1}$ = 2 (red), 2.5 (orange), 3 (yellow), 3.5 (green), 4.0 (light blue), 4.5 (dark blue), and 5 (pink). The x-axis corresponds to the $q_2$ value and the different panels to different break radius ($r_{\rm b}$). The solid lines adopt the non-unity ejection efficiencies discussed in Sections \ref{fejectpms} and \ref{fejectenc}, adopting for post-main sequence mass loss the results of Veras et al. (\citeyear{2014MNRAS.437.1127V}, namely, ${\it f_{\rm eject}^{\rm PMS}}({\it M_{*}})$ = 51.5\%, 74.7\%, 82.7\%, 85.3\%, 87.3\%, 87.8\%, 88.3\% and 89.0\%, for stellar masses of 1 $M_{\odot}$, 2 $M_{\odot}$, 3 $M_{\odot}$, 4$M_{\odot}$, 5 $M_{\odot}$, 6 $M_{\odot}$, 7 $M_{\odot}$, and 8 $M_{\odot}$, respectively), and adopting for the stellar encounters the results from Hanse et al. (\citeyear{2018MNRAS.473.5432H}, namely, ${\it f_{\rm eject}^{\rm enc}}({\it M_{*}})$ = 0.5).   The dot-dashed lines correspond to the assumption that all the bodies in the exo-Oort clouds are ejected. The black lines correspond to the cumulative number density of interstellar objects inferred from 1I/'Oumuamua's detection. 
}
\label{2s}
\end{figure}

The resulting contributions from 1--8 M$_{\odot}$ stars to the population of ejected exo-Oort cloud objects (calculated using Equations \ref{NPMS} and \ref{Nenc}) are listed in Table 2, assuming for the size distribution a single power law with index $q$ = 3.5. The results corresponding to the broken power law size distributions described in Section \ref{cumOu} are shown in Figure \ref{2s}, where in this case we are adding up the contributions of post-main sequence mass loss (Equation \ref{NPMS}) and stellar encounters (Equation \ref{Nenc}). The solid lines correspond to the ejection efficiencies discussed in Sections \ref{fejectpms} and \ref{fejectenc}, while the dot-dashed lines adopt the assumption that all the bodies in the exo-Oort clouds are ejected. The large differences among the  values shown in Table 2 and Figure \ref{2s} (also Figure \ref{2sl}) are due to the $\left({N_{r \geqslant R_1}\over N_{r \geqslant R_2}}\right)$ factor discussed above.

\section{Discussion}
\label{Discussion}

The hypothesis we want to test is whether 1I/'Oumuamua could be representative of a population of  exo-Oort cloud objects ejected at different stages of the star/planetary system evolution. To test this hypothesis, we compare the cumulative number densities calculated in Section \ref{Contribution} and the back-of-the-envelope estimate described in Section \ref{OverallUpperLimit},  summarized in Table 2 and Figure \ref{2s}, to that inferred from the detection of 1I/'Oumuamua. As mentioned in Section \ref{Comparison_back-of-envelope}, for the latter, we adopt the number density distribution derived by Do et al. (\citeyear{2018ApJ...855L..10D}) of $N_{\rm r \geqslant R}$ =  0.21 au$^{-3}$ $\sim$ 2$ \cdot~10^{15}$ pc$^{-3}$ (see Section \ref{Comparison_back-of-envelope} for a discussion of the uncertainties associated with this value). This observationally-inferred value derived by Do et al. (\citeyear{2018ApJ...855L..10D}) is shown as a black line in Figure \ref{2s} and is the last entry of Table 2.  

Table 2 shows that, when adopting the standard equilibrium power law size distribution with index $q$ = 3.5, $n(r) \propto r^{-3.5}$, the total contribution from 1--8 M$_{\odot}$ stars to the cumulative number density of ejected exo-OC objects with sizes equal or larger than 1I/'Oumuamua is of the order of $\sim$ 1.1$ \cdot~10^{13}$ pc$^{-3}$, when adopting the ejection efficiencies derived from Veras et al. (\citeyear{2014MNRAS.437.1127V}) and Hanse et al. (\citeyear{2018MNRAS.473.5432H}). As we calculated in Section \ref{Comparison_back-of-envelope}, if we were to assume all the objects in the exo-Oort clouds of these stars are ejected, this value would be $\sim$ 1.6$ \cdot~10^{13}$ pc$^{-3}$. Both values are about two orders of magnitude lower than the estimated value based on the detection of 1I/'Oumuamua. 

Figure \ref{2s} shows that, when adopting the broken, power law size distribution described in Section \ref{cumOu}, only the size distribution with the largest value of $q_1$ considered ($q_1$ = 5.0) would result in a cumulative number density within the uncertainty of that derived by Do et al. (\citeyear{2018ApJ...855L..10D}), with  $q_1$ = 4.5 fitting the lower end of the values estimated by other authors. 

The characteristic size distribution of exo-Oort cloud objects is of course unknown. To assess whether the required high value of $q_1$ is realistic, we take as a reference the size distribution derived for the population of minor bodies in the solar system. For the asteroid belt, Bottke et al. (\citeyear{2005Icar..175..111B}) estimated that its "primordial'' size distribution followed a broken power law with {\it q$_1$} $\approx$ 1.2, {\it q$_2$} $\approx$ 4.5, and {\it r$_b$} $\approx$ 50 km. For the Kuiper belt, Bernstein et al. (\citeyear{2004AJ....128.1364B}) calculated a broken power law with {\it q$_1$} = 2.9 and {\it q$_2$} $>$ 5.85 for the classical Kuiper belt, and {\it q$_1$} $<$ 2.8 and {\it q$_2$} = 4.3 for the excited Kuiper belt, with {\it r$_b$} $\leq$ 50 km in both cases, while Fraser and Kavelaars (\citeyear{2009AJ....137...72F}) found {\it q$_1$} = 1.9, {\it q$_2$} = 4.8, and {\it r$_b$} $\leq$ 25--47 km. For elliptic comets, the size distribution can be approximated by a single power law of $q \approx$ 2.9 for $r >$ 1.6 km or $q \approx$ 2.6, when including cometary near-earth objects (Lamy et al. \citeyear{2004come.book..223L}). Coagulation models that take into account the collisional evolution due to self-stirring find that the differential size distribution expected for Kuiper bel objects  follows a broken power law with
{\it n(r) $\propto$ r$^{-q_1}$} if {\it r $\leq$ r$_1$},  
{\it n(r) $\propto$ constant} if {\it r$_1$ $\leq$ r $<$ r$_0$}, and
{\it n(r) $\propto$ r$^{-q_2}$} if {\it r $\geq$ r$_0$}, 
where {\it r} is the planetesimal radius and $q_1$ $\approx$ 3.5 (resulting from the collisional cascade);  for a fragmentation parameter, {\it Q$_b$} $\gtrsim$ 10$^5$ erg/g,  $q_2$ $\approx$ 2.7--3.3, and  {\it r$_0$  $\approx$ r$_1$ $\approx$} 1 km;  while for {\it Q$_b$} $\lesssim$ 10$^3$ erg/g, $q_2$ $\approx$ 3.5--4, {\it r$_1$  $\approx$} 0.1 km, and {\it r$_0$  $\approx$} 10--20 km (see review in Kenyon et al. \citeyear{2008ssbn.book..293K}). 

Based on the above results, the required value of $q_1$ $\gtrsim$ 4.5 (more likely $q_1$ $\sim$ 5) might not be common in exo-Oort cloud objects (but, of course, this remains an open question), and this would yield to the conclusion that 1I/'Oumuamua is not representative of a background population of ejected exo-Oort cloud objects from 1--8 M$_{\odot}$ stars. 

\subsection{Is our estimate of the contribution of 1--8 M$_{\odot}$ stars an upper limit?}
\label{UpperLimit}

In Section \ref{Comparison_back-of-envelope}, we argued that our back-of-the-envelope calculation was an upper limit because it assumes that all stars contribute, irrespective of whether or not they are planet hosts, and that all the objects are ejected, irrespective of whether or not they form part of an exo-Oort cloud before being completely ejected from the system. The latter assumption is justified by the models from Veras et al. (\citeyear{2011MNRAS.417.2104V}) that find that, even if the exo-Oort cloud objects remain on stable orbits after the period of post-main sequence stellar mass loss, they may get ejected later on because the Galactic tide will be stronger relative to the star's gravity, shrinking the stable region as the orbits expand, or  because the influence from nearby stars will become stronger, leading to more ejections due to stellar encounters. 

Another consideration that makes our estimate an upper limit is dynamical heating. Ejected exo-Oort cloud objects will be scattered out of the Galactic disk due to passing stars, giant molecular clouds, or spiral arms, with the smaller objects being more strongly affected because of the conservation of momentum (Feng \& Jones \citeyear{2018ApJ...852L..27F}). Therefore, it is expected that the objects unbound from their parent stars early in the Galactic disk lifetime would no longer contribute to the population of interstellar objects in the Galactic disk due to dynamical heating. Because we are taking into account the contribution from all ejections, regardless of when the objects became unbound, our estimate is likely an upper limit. This strengthens our conclusion that it is unlikely that 1I/'Oumuamua is  representative of a background population of exo-Oort cloud objects ejected from 1--8 M$_{\odot}$ stars. 

\renewcommand\thetable{2}
\LongTables
\begin{deluxetable*}{ll}
\tablewidth{0pc}
\tablecaption{Cumulative number density of ejected exo-OC objects with sizes equal or larger than 1I/'Oumuamua\tablenotemark{a}.}
\tablehead{
\colhead{} &
\colhead{N$_{\rm exo-OC}$ (pc$^{-3}$)}}
\startdata
& \\
\multicolumn{2}{c}{\bf{Contribution from 1--8 M$_{\odot}$ stars}}\\
& \\
\multicolumn{2}{l}{\bf{Contribution from post-main sequence mass loss\tablenotemark{b}:}}\\
{$\int\limits_{1M_{\odot}}^{8M_{\odot}} \xi({\it M_{*}})\it f_{\rm PMS}({\it M_{*}})\left({\rm 0.16 \over \rm 2.3}\right)^{\rm -2.5}\rm 10^{12} \left({{\it M_{*}} \over {M_{\odot}}}\right){\it f_{\rm eject}^{\rm PMS}}({\it M_{*}})d{\it M_{*}}$}& 1.0$ \cdot~$10$^{13}$\\
& \\
\multicolumn{2}{l}{\bf{Contribution from stellar encounters\tablenotemark{c}:}}\\
{$\int\limits_{1M_{\odot}}^{8M_{\odot}} \xi({\it M_{*}})(\rm 1-\it f_{\rm PMS}({\it M_{*}}))\left({\rm 0.16 \over \rm 2.3}\right)^{\rm -2.5}\rm 10^{12} \left({{\it M_{*}} \over {M_{\odot}}}\right){\it f_{\rm eject}^{\rm enc}}({\it M_{*}})d{\it M_{*}}$}&1.3$ \cdot~$10$^{12}$\\
& \\
\multicolumn{2}{l}{\bf{Upper limit\tablenotemark{d}:}}\\
{$\int\limits_{1M_{\odot}}^{8M_{\odot}} \xi({\it M_{*}})\left({\rm 0.16 \over \rm 2.3}\right)^{\rm -2.5}\rm 10^{12} \left({{\it M_{*}} \over {M_{\odot}}}\right)d{\it M_{*}}$}& 1.6$ \cdot~$10$^{13}$\\
& \\
\hline
& \\
\multicolumn{2}{c}{\bf{Contribution from 0.08--1 M$_{\odot}$ stars}}\\
& \\
& \\\multicolumn{2}{l}{\bf{Contribution from stellar encounters\tablenotemark{c}}:}\\
{$\int\limits_{0.08M_{\odot}}^{1M_{\odot}} \xi({\it M_{*}})(\rm 1-\it f_{\rm PMS}({\it M_{*}}))\left({\rm 0.16 \over \rm 2.3}\right)^{\rm -2.5}\rm 10^{12} \left({{\it M_{*}} \over {M_{\odot}}}\right){\it f_{\rm eject}^{\rm enc}}({\it M_{*}})d{\it M_{*}}$} & 1.7$ \cdot~$10$^{13}$\\
& \\
\\
\hline
\\
\multicolumn{2}{l}{\bf{Inferred from 1I/'Oumuamua's detection:} }\\
 & 2.0$ \cdot~$10$^{15}$\\
\tablenotetext{a}{Assuming a single power law size distribution of index $q$ = 3.5.}
\tablenotetext{b}{Assuming for the ejection efficiency, ${\it f_{\rm eject}^{\rm PMS}}$, the values derived from the simulations by Veras et al. (\citeyear{2014MNRAS.437.1127V}; see discussion in Section \ref{fejectpms}).}
\tablenotetext{c}{Assuming an  ejection efficiency ${\it f_{\rm eject}^{\rm enc}}$ = 0.5, based on dynamical simulations by Hanse et al. (\citeyear{2018MNRAS.473.5432H}) for the Sun's Oort cloud (see discussion in Section \ref{fejectenc}).}
\tablenotetext{d}{See discussion in Section \ref{OverallUpperLimit}.}
\end{deluxetable*}

\subsection{Contribution from lower-mass stars due to stellar encounters}
\label{LowMassStars}

As mentioned in Section \ref{Clearing_PMS}, post-main sequence mass loss of stars with masses $<$ 1 M$_{\odot}$ do not contribute to the population of ejected exo-Oort cloud bodies because their mass loss is not strong enough to unbind these objects. However, given their high number density, stellar encounters of low-mass stars could contribute significantly to the population of ejected exo-Oort cloud bodies if exo-Oort clouds around these stars were to be common. In this regard, Wyatt et al. (\citeyear{2017MNRAS.464.3385W}) argues that these Oort clouds could be populated by objects scattered by low mass planets like the Earth located at 5 au, with the low mass of the star facilitating the scattering of objects encountering the planets. It remains unknown, however, how frequent these exo-Oort clouds may be around low-mass stars, with the higher frequency of planets around these stars maybe facilitating their formation. 

\begin{figure}
\begin{center}
\includegraphics[width=8cm]{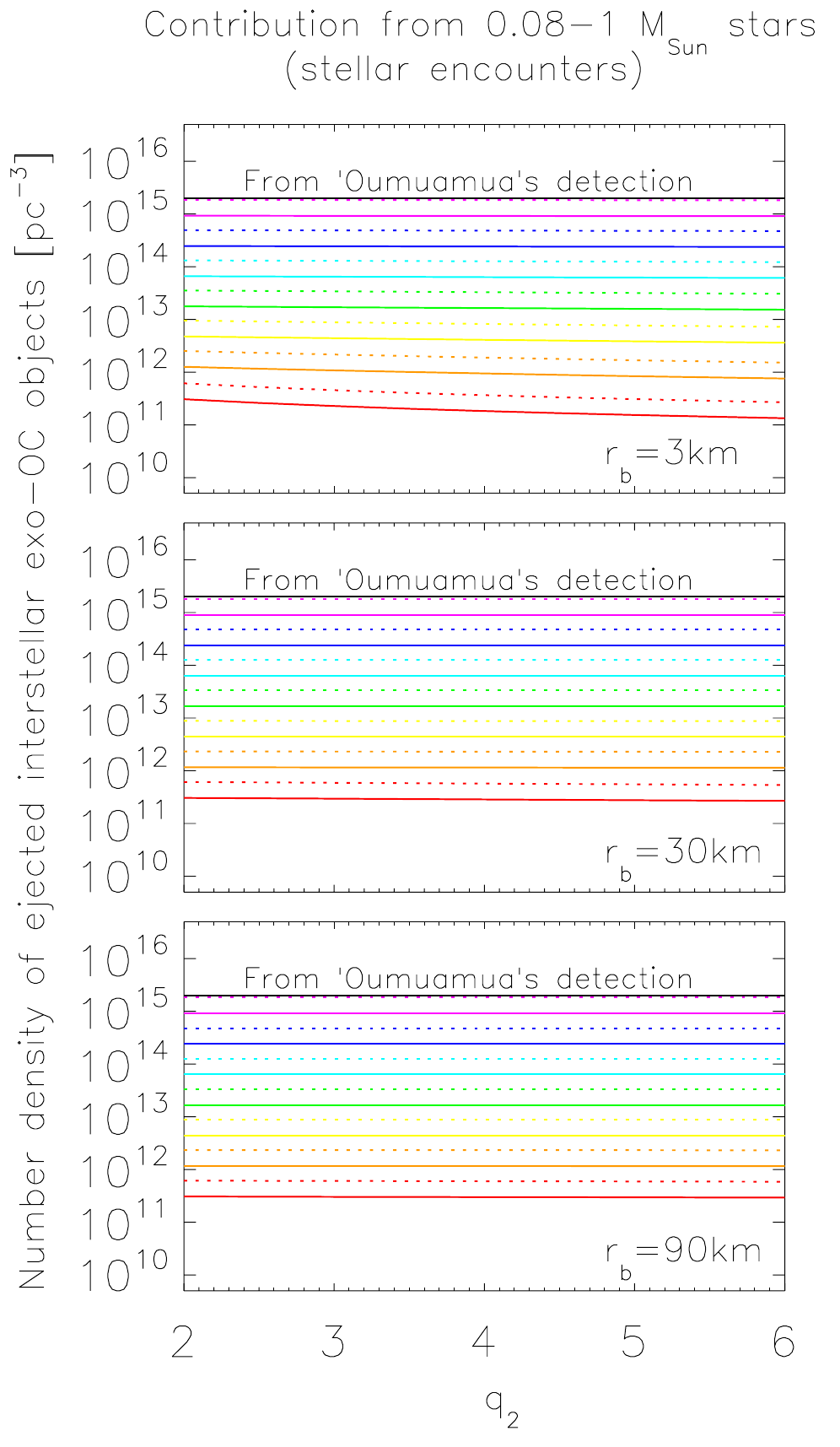}
\end{center}
\caption{Same as Figure \ref{2s} but showing the contribution due to stellar encounters of 0.08--1~$M_{\odot}$ stars. Post-main sequence mass loss is not expected to contribute to the population of ejected objects in this stellar mass range.
}
\label{2sl}
\end{figure}

Assuming the scenario that they are frequent, we estimate their contribution to the cumulative number density of ejected exo-OC objects with sizes equal or larger than 1I/'Oumuamua by expanding the integral in Equation \ref{Nenc} to include stars with masses 0.08 $\leqslant$ M$_{*}$ $<$ 1 M$_{\odot}$. The results are listed in Table 2 and in Figure \ref{2sl}.  As indicated above, the large differences among the values shown are due to the $\left({N_{r \geqslant R_1}\over N_{r \geqslant R_2}}\right)$ factor discussed above. As Table 2 and in Figure \ref{2sl} illustrate, the contribution from stellar encounters of low mass stars, being about twice that of the contribution from post-main sequence mass loss and stellar encounters of 1--8~$M_{\odot}$ stars, could still not account for the inferred cumulative number density of interstellar objects based on 1I/'Oumuamua's detection, unless the $q_1$ index of the size distribution is large ($q_1 \sim$ 5.0) and we assume all the objects of these exo-Oort clouds are ejected (shown as dotted lines in Figure \ref{2sl}).  

But in the case of low mass stars, there is another consideration to be taken in to account and is the low incoming velocity of I/'Oumuamua, within 3--10 \kms~of the LSR (Gaidos et al. \citeyear{2017RNAAS...1a..13G}; Mamajek \citeyear{2017RNAAS...1a..21M}, Do et al. \citeyear{2018ApJ...855L..10D}). This makes a low-mass stellar origin less likely as these stars tend to have larger dispersion velocities because they tend to be older and because low-mass objects are more affected by dynamical heating (Dehnen \& Binney \citeyear{1998MNRAS.298..387D}). This kinematic constraint is also relevant when considering the contribution of  1--8~$M_{\odot}$ stars, as it includes that of old stars with higher dispersion velocities (as mentioned in Section \ref{UpperLimit}). 

\subsection{Could there be other discriminating measurements regarding 1I/'Oumuamua;s origin?}

\subsubsection{Flux of micrometeorites and meteorites on Earth}
\label{Met}
In Moro-Mart\'{\i}n (\citeyear{2018ApJ...866..131M}), where we explored a potential protoplanetary disk origin for 1I/'Oumuamua, we addressed whether the flux of meteorites and micrometeorites expected on Earth could be a discriminating measurement regarding the origin of interstellar interlopers, as other authors had suggested that one of these interstellar objects may already be part of the collected meteorite samples (Gaidos (\citeyear{2018MNRAS.477.5692G}). We did this by comparing the observed fluxes for meteorites and micrometeorites on Earth to those inferred from 1I/'Oumuamua's detection, and to those expected from a planetesimal disk origin. We found that the observed fluxes on Earth are 8--10 orders of magnitude larger than the latter, indicating that these fluxes cannot be used as a discriminating factor (because the contribution of the extrasolar impactors to these two fluxes would be negligible; see Table 2 in Moro-Mart\'{\i}n \citeyear{2018ApJ...866..131M}). We also concluded that it is unlikely that the collected meteorite samples include an interstellar meteorite.  Given that the cumulative number density of interstellar objects expected from an exo-Oort cloud origin (listed in Table 2 in this paper) is only about one order of magnitude larger than that expected from a protoplanetary disk origin, the conclusion remains the same as in the protoplanetary disk case: the observed fluxes on Earth of meteorites and micrometeorites cannot be used as discriminating measurement regarding a potential exo-Oort cloud origin. 

\subsubsection{Number density of free-floating, planetary-mass objects from gravitational microlensing surveys}
\label{FreeFloaters}

Another potential discriminating measurement could be the number density and mass distribution of the free-floating, planetary-mass objects, as these objects could have been released as the result of the same dynamical processes considered in this paper for the ejection of exo-Oort cloud objects. A large gravitational microlensing survey published recently by Mr{\'o}z et al. (\citeyear{2017Natur.548..183M}), studying $\sim$ 2600 events, found that the frequency of free-floating, Jupiter-mass planets is about 0.25 with a 95\% confidence level (although these planets could also be in wide orbits). They also found six very short events that could be due to Earth or super-Earth-mass planets that are either free-floating or on wide orbits. They argued that the small number of these very short event detections and their uncertain origin do not allow to infer a mass distribution for the free-floating population. However, they pointed out that if we were to assume that 5M$_{\oplus}$ mass planets are five times more frequent than main sequence stars, the expected number of these very short microlensing events would be 2.2 (compared to the six observed). 

Assume for a moment we adopt that frequency as a rough estimate for free-floating planets of $\sim$ 2R$_{\oplus}$ radius (corresponding to $\sim$ 5M$_{\oplus}$ for H$_2$O composition). If we were to assume that there is a characteristic size distribution that connects the super-Earth planet population to the smaller bodies, we could in principle compare that rough estimate to the cumulative number density of super-Earths with sizes larger than  $\sim$ 2R$_{\oplus}$ that we would expect based on both, the cumulative number density of objects larger than 1I/'Oumuamua inferred from its detection ($N_{\rm r \geqslant R}$ $\sim$ 2$ \cdot~10^{15}$ pc$^{-3}$), and that expected from an exo-Oort cloud origin ($N_{\rm r \geqslant R}$ $\sim$ 1$ \cdot~10^{13}$ pc$^{-3}$, see Table 2). 

If we were to adopt for the characteristic size distribution an equilibrium power law with index $q$ = 3.5, $n(r) \propto r^{-3.5}$, and a maximum object radius of 3R$_{\oplus}$ (encompassing the free-floating super-Earths), we would get that the expected cumulative number density of super-Earths  $\gtrsim$ 5M$_{\oplus}$ would be $\sim$1--200 pc$^{-3}$, which is $\sim$ 1--200 $\times$ the number density of main sequence stars, compared to the rough estimate of 5 $\times$ more frequent derived from the microlensing survey mentioned above. If instead we were to adopt a size distribution based on coagulation and collisional models of the small-body population in the solar system Kuiper belt, as the one derived by Schlichting et al. (\citeyear{2013AJ....146...36S}) [described in Equation (2) in Moro-Mart\'{\i}n (\citeyear{2018ApJ...866..131M})], but assuming it also applies to objects as large as 3R$_{\oplus}$, we would get that the expected cumulative number density of super-Earths  $\gtrsim$ 5M$_{\oplus}$ would be $\sim$ 0.05--10 pc$^{-3}$, which is $\sim$ 0.05--10 $\times$ the number density of main sequence stars [this latter calculation is done using Equations (27)--(33) in Moro-Mart\'{\i}n (\citeyear{2018ApJ...866..131M})], compared again to the rough estimate of 5 $\times$ more frequent from the microlensing survey. These rough estimates indicate that, unlike the case of the micrometeorite and meteorite fluxes where we found many order of magnitude differences, for the adopted equilibrium size distribution, these two independent estimates of the population of interstellar objects (from the detection of interlopers and from microlensing surveys) could be of the same order, leaving open the possibility of a common origin. 

The main caveat of the above argument (in which we assumed 10$^{12} \left({{\it M_{*}} \over {M_{\odot}}}\right)$ objects per star and adopted an equilibrium size distribution) is that the number of objects in the exo-Oort clouds (or total mass distribution) and the size distribution of the exo-Oort cloud bodies are unknown and, regarding the latter, there might not be a characteristic one that links the small-body population to the super-Earths. WFIRST will be able to study the mass distribution of the free floating planets down to Mars-sized objects (Spergel et al. \citeyear{2015arXiv150303757S}). However, it is unlikely that these objects are part of a collisional population that encompasses 1I/'Oumuamua's-sized objects (or any expected interloper). The detection of the latter, unfortunately, could remain beyond the capabilities of microlensing surveys, even beyond WFIRST. This yields to two conclusions: (1) It seems unlikely that the microlensing surveys could be used as a discriminating measurement to address the origin of the interlopers. (2) The detection of interlopers may be one of the few observational constraints to shed light on the low-end of the mass distribution of the free-floating population, with the caveat that, as we have seen in this study and in Moro-Mart\'{\i}n (\citeyear{2018ApJ...866..131M}), in the case of 1I/'Oumuamua, it might not be appropriate to assume this object is representative of an isotropic background population of interlopers, which makes the derivation of its number density very challenging.

\section{Conclusion}
\label{Conclusion}

In this paper, we test the hypothesis that 1I/'Oumuamua, the first interloper to have ever been detected, is representative of a background population of interstellar objects ejected from exo-Oort clouds. We are interested in this potential origin  because dynamical models show that, over the lifetime of their parent stars, these weakly bound objects, of predominantly icy composition, are subjected to ejection due to Galactic tides, post-main sequence mass loss, and encounters with other stars or with giant molecular clouds, contributing to the population of free-floating material. 

To test this hypothesis, we compare the cumulative number density of interstellar objects expected from an exo-Oort cloud origin to that inferred from the detection of 1I/'Oumuamua. For the exo-Oort clouds, we use the solar system's Oort cloud as a reference and scale the inner and outer edge to the Hill radius of the parent star in the Galactic potential. We also assume that the number of bodies scales with the mass of the star.  Based on the work by Veras et al. (\citeyear{2014MNRAS.437.1127V}), we then estimate the regimes of influence of the different exo-Oort cloud clearing processes mentioned above and conclude that, given our exo-Oort cloud assumptions, the two processes expected to dominate the ejection are: (1) post-main sequence mass loss, for the stars in the 1--8 M$_{\odot}$ mass range that have reached this stage of stellar evolution; and (2) stellar encounters of 0.08--8 M$_{\odot}$ stars that are still on the main sequence. Adopting dynamical modeling results from other authors regarding the efficiency of ejection of these two processes, we estimate their expected contribution to the population of ejected exo-Oort cloud objects. We take into account dependencies with the stellar mass, Galactocentric distance, and evolutionary state, and consider a wide range of possible size distributions for the ejected objects. 

Regarding the contribution of exo-Oort clouds around 1--8 M$_{\odot}$ stars to the interstellar population of unbound objects (due to post-main sequence mass loss and stellar encounters), we find that this origin is unlikely because the expected cumulative number density of ejected objects from this population is significantly lower than the estimated value based on the detection of 1I/'Oumuamua. Regarding the contribution of exo-Oort clouds around low-mass (0.08--1 M$_{\odot}$) stars due to stellar encounters, we reach the same conclusion deeming this origin also unlikely. Our conclusion is strengthened by the consideration that our estimate is likely an upper limit. 

We find that the flux of micrometeorites and meteorites on Earth cannot be used as a discriminating measurement regarding a potential exo-Oort cloud origin because the observed fluxes are 8--10 orders of magnitude larger than expected (indicating exo-Oort clouds would make a negligible contribution). The other potential discriminating measurement we explore is the number density of free-floating, planetary-mass objects derived from gravitational microlensing surveys. We conclude that, given the mass limitations of the surveys and the resulting uncertainty in the mass distribution of the free-floating material, it seems unlikely that the results from these surveys could be used as a discriminating measurement to address the origin of the interlopers. The detection of interlopers may be one of the few observational constraints of the small end of this free-floating population, with the caveat that, as we conclude here and in Moro-Mart\'{\i}n (\citeyear{2018ApJ...866..131M}), in the case of 1I/'Oumuamua, it might not be appropriate to assume this object is representative of an isotropic background population, which makes the derivation of a number density very challenging. 

A. M.-M. thanks the anonymous referee for helpful suggestions that have significantly improved the clarity of the manuscript and Alycia J. Weinberger for her feedback in the calculation of the low-mass star contribution.

\end{document}